\documentclass[11pt,twocolumn,showpacs,floatfix,superscriptaddress,nofootinbib,amssymb,amsmath,aps,prc]{revtex4-2}

\usepackage{CJK} 
\usepackage[dvips]{graphicx}
\usepackage{epstopdf} 
\usepackage{latexsym,amssymb,amsmath,epsfig,bm,times,psfrag,subfig}
\usepackage{color}
\usepackage{mathptmx}                
\usepackage{dcolumn}                 
\usepackage[bookmarksnumbered,bookmarksopen,colorlinks,citecolor=blue,linkcolor=blue]{hyperref}
\renewcommand{\part}{{\rm part}}

\newcommand{\be}{\begin{equation}}
\newcommand{\ee}{\end{equation}}
\newcommand{\bear}{\begin{eqnarray}}
\newcommand{\eear}{\end{eqnarray}}
\newcommand{\ba}{\begin{array}}
\newcommand{\ea}{\end{array}}

\begin{document}
	\begin{CJK*}{UTF8}{gbsn}

	\title{Studying Effects of  Evolution on Charge Separation in Small Collision System \\ with  A MultiPhase Transport model}

	\author{Yi Xu (许易)}
    \affiliation{School of Physics, Huazhong University of Science and Technology, Wuhan 430074, China}

	\author{Chen Gao (高晨)}
	\affiliation{School of Physics, Huazhong University of Science and Technology, Wuhan 430074, China}

	\author{Shi-Xue Zhang (张世学)}
    \affiliation{School of Physics, Huazhong University of Science and Technology, Wuhan 430074, China}
	
	\author{Wei-Tian Deng (邓维天)}
	\thanks{The corresponding author}
	\email{dengwt@hust.edu.cn}
	\affiliation{School of Physics, Huazhong University of Science and Technology, Wuhan 430074, China}


\begin{abstract}
In relativistic high energy heavy-ion collisions,  Chiral Magnetic Effect (CME) could produce a charge separation in QGP. The charge separation could survive into final hadron system during evolution, observed as correlator $\Delta\gamma$. This physics procedure could also occurs in small collision system if CME  emerge in the QGP droplet. In this paper, we study  the effects of QGP droplet evolution on  charge separation in small collision system, based on AMPT model.  Our calculation indicates that, with given initial charge separation, the effects of parton level evolution and hadron level evolution weaken the charge separation indeed, but there is still enough signals could remained. Furthermore, we found that, the contribution of background to $\Delta\gamma$ is negligible in small collision system.  So, we propose the small collision system being a good place to confirm contribution of CME.
\end{abstract} 
\maketitle
\end{CJK*}

\section{Introduction\label{sec:intro}}

In relativistic heavy-ion collisions,   the extremely high temperature needed for quantum chromodynamics (QCD) phase transition is satisfied,  quarks and gluons are deconfined, forming a kind of new matter called quark-gluon plasma (QGP). In off-center heavy ion collisions, an extremely strong transient electric and magnetic fields are also generated \cite{Rafelski:1975rf,Skokov:2009qp,Bzdak:2011yy,Voronyuk:2011jd,Deng:2012pc,Deng:2014uja,Inghirami:2016iru,Yan:2021zjc}. 
 On the other hand, people has realized that there may be a parity symmetry ($\mathcal{P}$) or charge conjugation and parity symmetry ($\mathcal{CP}$) violation effects due to quantum anomalous fluctuation of  QCD vacuum ~\cite{Kharzeev:1998kz,Kharzeev:2004ey,Kharzeev:2007jp,Fukushima:2008xe,Kharzeev:2015znc,Liu:2020ymh}. This effect could cause a charge separation in QGP due to the coupling of $\mathcal{CP}$ violation and  strong magnetic fields, called as chiral magnetic effect (CME) ~\cite{Kharzeev:2015znc, Kharzeev:2009fn, Fukushima:2010vw,Basar:2010zd}.
 
As the local charge separation induced by CME can't be observed directly in experiment,  the  three-point azimuthal correlator is introduced~\cite{Voloshin:2004th}  to measure the charge imbalance.
 \begin{equation}
 		\gamma_{\alpha\beta}=< \cos (\Phi_{\alpha}+\Phi_{\beta}-2\Phi_{RP})>
 \end{equation}
 Where $\Phi_{\alpha}$ and $\Phi_{\beta}$  are azimuthal angles of charged particle $\alpha$ and $\beta$,  $\Phi_{RP}$  is the azimuthal  direction of  reaction plane .
 Also, in order to erase the charge-independent backgrounds like local momentum conservation, the difference of correlator 
 \begin{equation}
 	 \Delta \gamma = \gamma_{OS} - \gamma_{SS}
 	 \label{Eq-Delta-gamma1}
 \end{equation}
is introduced ~\cite{Kharzeev:2015znc}, where  $\gamma_{OS}$ represents for opposite charge particle pair, $\gamma_{SS}$ represents for same charge particle pair.  Beside the contribution from CME to charge correlator $\Delta \gamma$, there is other QCD background effects that can contribute to  $\Delta \gamma$ too. These effects are related to QGP flow $v_2$~\cite{Xu:2017zcn,STAR:2013zgu,Wang:2016iov,Ajitanand:2010rc}.  So, the correlator observed on experiment include two contributions indeed
 \begin{equation}
	\Delta\gamma=\Delta\gamma_\mathrm{CME} +\Delta\gamma_\mathrm{background}.
	\label{Eq-Delta-gamma}
\end{equation} 
The contribution of CME into correlator $\Delta\gamma$ is proportional to the angular difference between $\phi_B$ and $\phi_{RP}$, expressed as~\cite{Bloczynski:2013mca}
\begin{equation}
	\Delta\gamma_\mathrm{CME} \propto B^2\cdot\cos [2(\Phi_B-\Phi_{RP})],
	\label{Eq-Delta-gamma-PhiRP}
\end{equation} 
where $\Phi_B$ is the azimuthal angle of magnetic field.  
In order to confirm and clarify the contribution $\Delta\gamma_\mathrm{CME}$, an isobaric collision was proposed~\cite{Deng:2016knn}. While the experiment data~\cite{STAR:2021mii} recently indicated that these background contributions $\Delta\gamma_\mathrm{background}$ are quite large. So it is still a difficult  task to confirm the contribution from CME in heavy ion collision experiment.

In our previous work, we proposed a new method to check and clarify the contribution of CME in small collision system~\cite{Zhang:2021jrc,Wu:2024vcd}.  However, people used to think of there should be no charge separation caused by CME effect in small systems collision for two reasons. The first one is people believed the absence of azimuthal correlation between $\Phi_B$ and $\Phi_2$ \cite{Zhao:2017rpf, Belmont:2016oqp}, so the observation $\Delta\gamma_\mathrm{CME}$ could not be remained after average over many events .  The second reason is related with the small scale of time and space of QGP droplet in small collision system. People assume the initial charge separation, if there is, could not remained into the final hadron system after whole evolution process in small collision.

As already discussed about the first one  in our previous work ~\cite{Zhang:2021jrc},  we found that the orientation of  E-M field produced in $p+A$ has a non-vanished correlation with $\phi_{RP}$. Furthermore,we found that  in the small collision system evolving a polarized proton $p^{\uparrow} + A$, there are two particular colliding schemes in which the  correlator $\Delta\gamma_\mathrm{CME}$ has a  significant dependence on the angle between reaction plane and polarization direction~\cite{Wu:2024vcd}.  So the initial charge separation could be produced by local strong violation in the small collision system $p+A$ and $p^{\uparrow} + A$.  

In this paper, we concentrate on the discussion about the second reason, i.e.  the effects of evolution with an given initial charge separation in small collision system. We calculate the survived charge separation described as $\Delta \gamma$ after parton level evolution and hadron level evolution.  The content of this article is as follows:  In the second part, we introduce our calculation framework briefly . The  results of calculations are given in the third part.  At last, we give a summary about our calculation. 
 
\section{Calculation Framework\label{sec:method}}

In this work,  A MultiPhase Transport (AMPT) model~\cite{Zhang:1999bd, Lin:2004en} is employed to perform the simulation of $p+Au$ collisions with RHIC energy with $\sqrt{s}=200$ GeV.  

In the AMPT model, an event of heavy ion collision ($A+A$ or $p+A$) is divided into four stages:  to produce initial condition of parton level,  the evolution of parton level,  hadronization of partons into hadrons, and the evolution of hadron level. In the first stage, the Heavy Ion Jet INteraction Generator (HIJING)~\cite{Wang:1991hta, Gyulassy:1994ew} model is used to simulate heavy ion collision. The wounded nucleons and produced minijets are treated as excited strings attached by  partons.  About  stage two and stage three, there are two distinct versions of AMPT model available: the default version, and the string melting version. In the default version, these attached partons will suffer a parton level evolution described as Zhang parton cascade (ZPC) process ~\cite{Zhang:1997ej}. After that, the partons are recombined with excited strings. Then hadronization occurs through the Lund string fragmentation model~\cite{Andersson:1986gw, Nilsson-Almqvist:1986ast, Zhang:2000nc} to generate hadrons. In the string melting version, all excited strings are melted into partons. All of these partons  suffer ZPC paton cascade in the second stage. Then hadrons are produced through a quark recombination model~\cite{Lin:2002gc, Lin:2001zk} in the third stage. In the stage four, after all partons are hadronized into hadrons, A Relativistic Transport (ART)~\cite{Li:1995pra} model is used to describe the hadron level evolution.  

In our calculation, the string melting version of AMPT model is used to simulate the evolution of QGP droplet which is produced in high energy $p+Au$ collision.  the projectile is set as proton flying with $E=100$ GeV along $+z$ direction, while target is $Au$ flying with $E=100$ GeV per nucleon along $-z$ direction.  In order to concentrate on the effects of evolution on charge separation, we set the impact parameter vector $\mathrm{b}$ lies on $x$ coordinate pointing from $p$ to $Au$. So the azimuthal direction of reaction plane $\phi_{RP}$ is 0 in the following calculation. And direction of  magnetic field is on $-y$ direction on average. In ZPC parton evolution step, a 10mb parton cross section is set. 

Since there is no  CME mechanism in current AMPT models, we put a new stage into AMPT model after string melting to simulate the initial charge separation along magnetic fields induced by CME. Following the method in work~\cite{Ma:2011uma}, a few percentage of quarks are selected randomly,  their $y$-component of momentum are interchanged to form an initial charge separation. In details,  the $p_y$ of upward-moving $\bar{u}$ and  downward-moving $u$ are interchanged, and the $p_y$ of  upward-moving $d$ and downward-moving $\bar{d}$ are interchanged correspondingly.  After this stage is accomplished, all partons enter ZPC evolution stage to suffer the latter evolution. 

To determine the reasonable percentage of  interchanged quarks, we performed  simulation of $Au+Au$ collisions firstly to fit the  observed experiment data of $\gamma_{\alpha\beta}$ ~\cite{STAR:2009wot,STAR:2009tro}.
\begin{figure}
	\centering
	\includegraphics[width=0.85\linewidth,clip]{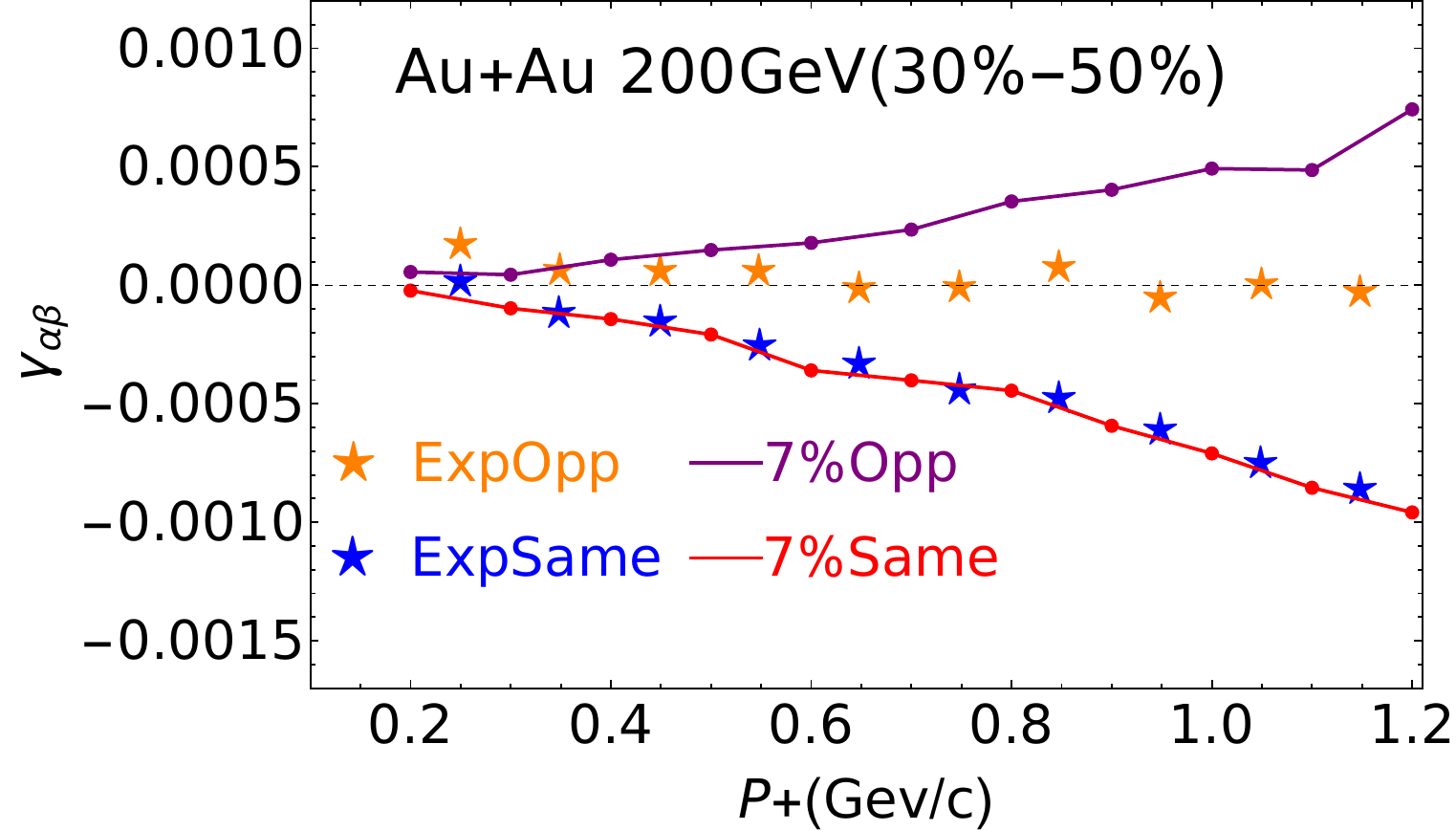}
	\caption{$\gamma_{\alpha\beta}$ as a function of $P_+$ for $Au+Au$ collisions with  30\%-50\% centrality.  Percentage of 7\% initial charge separation is set in AMPT calculation to fit these data. }
	\label{fig:AA}
\end{figure}
Shown in Fig.~\ref{fig:AA} is the $ \gamma_{\alpha\beta}$ of $Au+Au$ collisions in centrality of  30\% to 50\% as function of $P_+$, which is defined as
\begin{equation}
	P_+=\frac{P_{T,\alpha}+P_{T,\beta}}{2},
	\label{Eq-Pplus}
\end{equation}
where $P_{T,\alpha}$ and $P_{T,\beta}$ are transverse momentum of particle $\alpha$ and $\beta$.
After a global fitting, we found that the percentage of 7\% for initial charge separation can fit data of $\gamma_{SS}$ very well. On the same time,  we can see the failure of fitting data of $\gamma_{OS}$ like what in~\cite{Ma:2011uma}.  The asymmetry of $\gamma_{SS}$ and $\gamma_{OS}$ maybe caused by some charge-independent background effects existing in heavy ion collision.  

In our following calculation, this percentage of 7\%  is used for $p+Au$ collision, since we have found that the value of correlator $\Delta\gamma_\mathrm{CME}$ in $p+Au$ collisions is the same order of that in $Au+Au$ collision~\cite{Zhang:2021jrc,Wu:2024vcd}. And, the main goal of this work is to study the effects of evolution on charge separation,  to check how much initial charge separation  could remained to final hadron stage. So the percentage of 7\% could be considered as a reasonable upper limit of CME effect in $p+Au$ collision.

\section{Results \label{sec:Result}}

In this section, we calculate the charge correlator $\gamma_{\alpha\beta}$   in  $p+Au$ collisions with an initial charge separation. We show the results as function of impact parameter $b$, kinetic variable $P_+$ and $\Delta\eta$ respectively  .  On each data point in the following figures,  the statistic  error  is shown as an error bar. This error is defined as standard  deviation $\sigma=(\sqrt{\sum{{({s_i}-{<s>})}^2}})/N$.

Firstly, let's check  the correlator $\gamma_{\alpha\beta}$ as function of impact parameter $b$. 

\begin{figure}
	\centering
	\includegraphics[width=0.85\linewidth,clip]{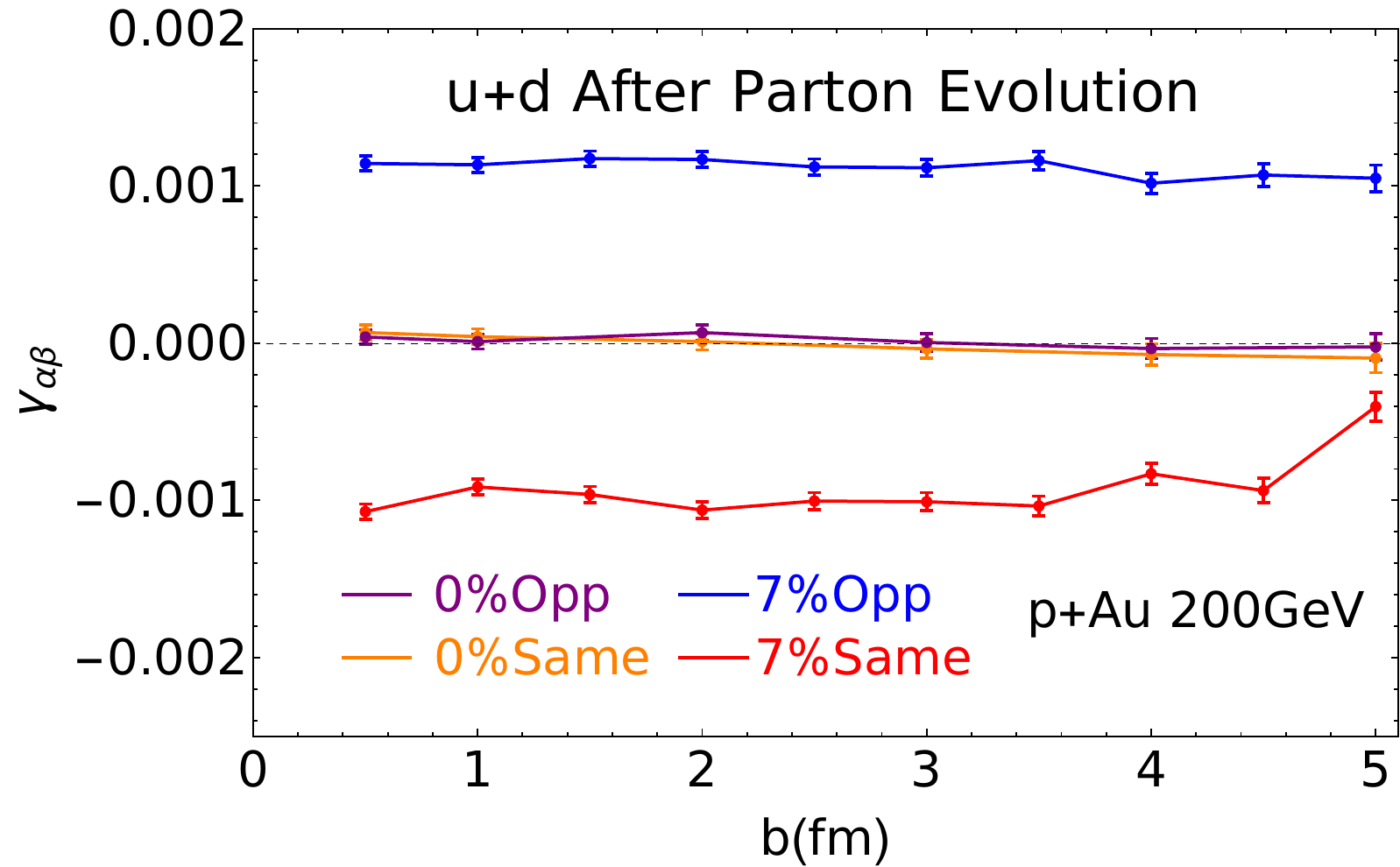}
	\caption{$\gamma_{\alpha\beta}$ as a function of impact parameter $b$ for $p+Au$ collisions  after ZPC parton level evolution. Percentage 7\%  and 0\%  of initial charge separation are shown for comparison. }
	\label{fig:zpcmostcentral}
\end{figure}

In this calculation, we set the interchange percentage 7\% as  $b$-independent, so the initial value of correlator $\gamma_{\alpha\beta}$ is almost $b$-independent. Their values  for paton system after quarks interchange stage are $\gamma_{OS} {\approx} 1.58 {\times} 10^{-3}$ and $\gamma_{SS} {\approx} -1.38 {\times} 10^{-3}$ correspondingly. 
After parton evolution, the value of  $\gamma_{\alpha\beta}$ should be weakened as we expected, since the azimuthal correlation between quarks would be disrupted by cascade process.  Shown in Fig.~\ref{fig:zpcmostcentral} ,  the value of $\gamma_{\alpha\beta}$ decrease to about $\pm 0.001$ indeed, which are 75\% of initial values for both  $\gamma_{OS}$ and  $\gamma_{SS}$ . This means that the effects of parton evolution on charge separation is charge-independent. Also, we can see their values are almost centrality independent except for that at large impact parameter $b>4$ fm.  The $\gamma_{SS}$ shows a depression in this region, this is only because the number of partons decreases in that off-central collisions, then the number of quarks interchanged maybe much less than small and middle impact parameter even with the same percentage. 

\begin{figure}[hb]
	\centering
	\includegraphics[width=0.85\linewidth,clip]{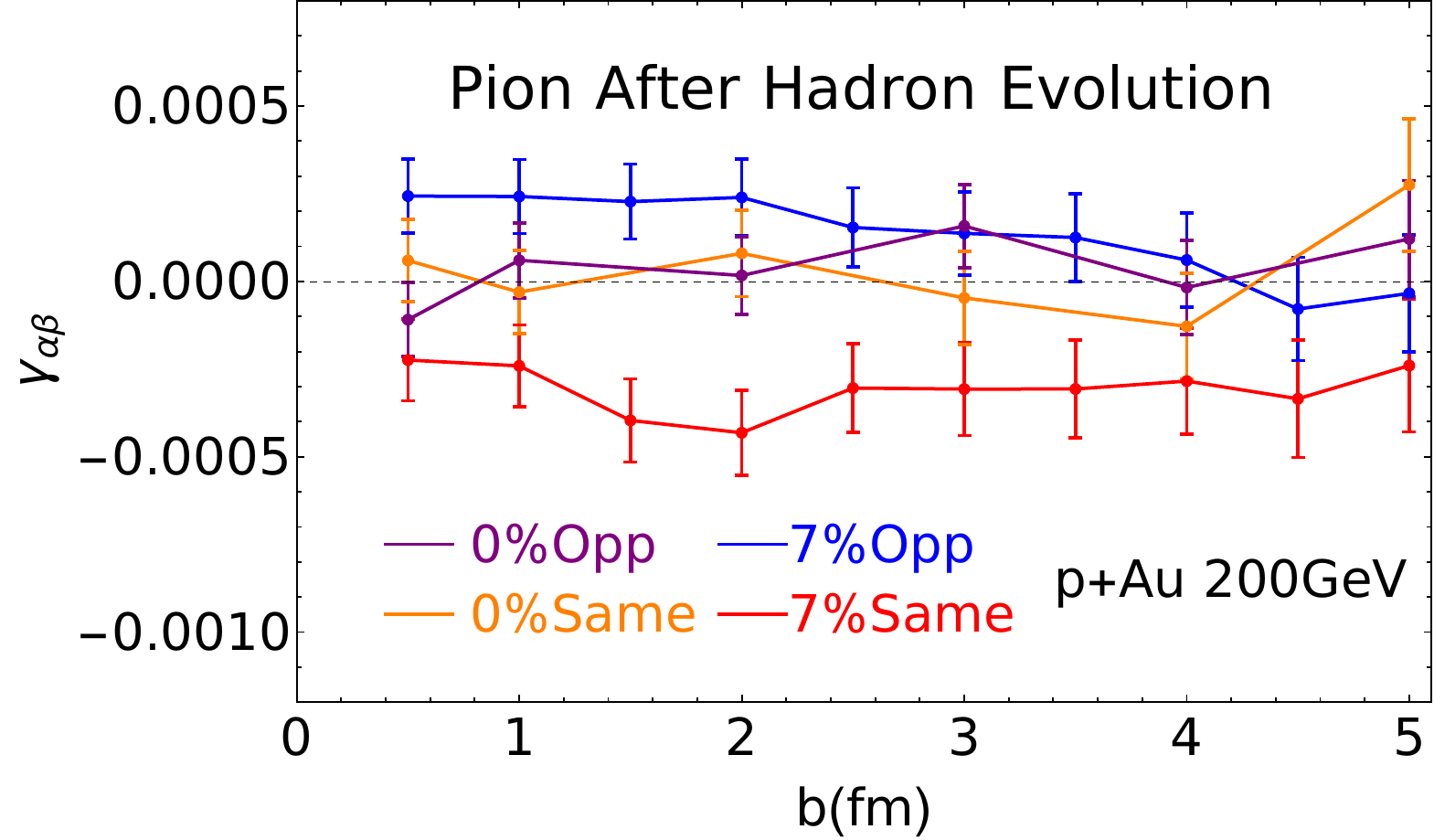}
	\caption{$\gamma_{\alpha\beta}$ as a function of impact parameter $b$ for $p+Au$ collisions  after hadron level evolution.  Results of  Percentage 7\%  and 0\%  of initial charge separation are shown for comparison. }
	\label{fig:amptmostcentral}
\end{figure}

After hadronization and hadron level evolution, the  correlator $\gamma_{\alpha\beta}$ should be weakened furthermore. Shown in Fig.~\ref{fig:amptmostcentral}, we can see about 40\%  charge separation survived into final hadron system indeed as we expected.  Comparing results of $\gamma_{OS}$ and $\gamma_{SS}$ is interesting. We can see that the value of  $\gamma_{OS}$ depressed more than $\gamma_{SS}$ obviously, especially at middle and large impact parameter.  

Furthermore , we give a comparison between results with interchange percentage 7\% and 0\%. The latter one means there is no any CME mechanism in our calculation.  Fig.~\ref{fig:zpcmostcentral}  shows that the results with 0\% after parton level evolution are still zero at each impact parameter. And in Fig.~\ref{fig:amptmostcentral}, the zero results remained into final hadron system within statistic error-bar.  These results are quite different with what in $Au+Au$ collision observed by paper~\cite{Ma:2011uma}. In  $Au+Au$ collision, a non-zero  $\gamma_{\alpha\beta}$  is got after parton level evolution with 0\%  interchange percentage.  And this non-zero results are expanded after hadron level evolution.    
These tells us that background contribution $\Delta\gamma_\mathrm{background}$ in $Au+Au$ is quite large while that in $p+Au$ is negligible.  These results hint us the observed significant correlator $\Delta\gamma$ in CMS experiment ~\cite{CMS:2016wfo}  could come from CME mainly.

Then, we checked  the correlator $\gamma_{\alpha\beta}$ as function of kinematic variable $P_+$ for both parton level and hadron level. $P_+$ is defined as Eq.~\ref{Eq-Pplus}, which means the sum of transverse momentum of  the pair particles. In this calculation, impact parameter $b=0.5$ fm is set. 

\begin{figure}[hb]
	\centering
	\includegraphics[width=0.85\linewidth,clip]{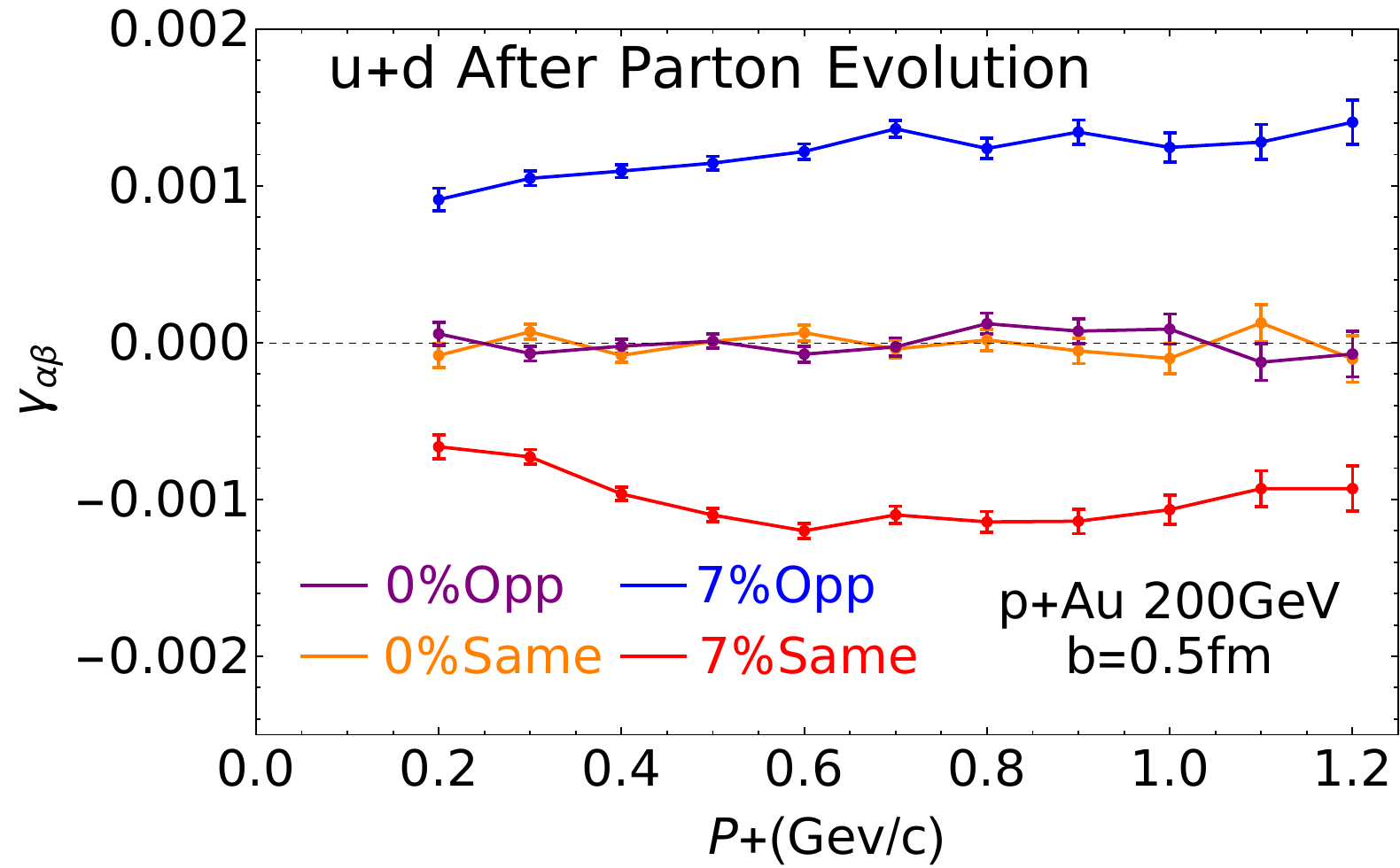}
	\caption{$\gamma_{\alpha\beta}$ as a function of  $P_+$ for $p+Au$ collisions  after ZPC parton level evolution.  Results of  Percentage 7\%  and 0\%  of initial charge separation are shown for comparison. } 
	\label{fig:zpcpplus}
\end{figure}

In Fig.~\ref{fig:zpcpplus},  the results of $\gamma_{\alpha\beta}$ with 7\% initial charge separation after parton level evolution are shown. We can see their values are almost $\pm 0.001$ again with a slight expanding with increase of $P_+ $. The values of   $\gamma_{OS}$ and $\gamma_{SS}$  are symmetric approximately in the whole range of $P_+$.   And in the range of $P_+ >0.6$ GeV,  the values of both $\gamma_{OS}$ and $\gamma_{SS}$ get a  saturation to $~0.0012$. This phenomenon tell us the effects of parton evolution on charge separation has only a weak dependence on kinematic viable  within AMPT model. 

\begin{figure}[ht]
	\centering
	\includegraphics[width=0.85\linewidth,clip]{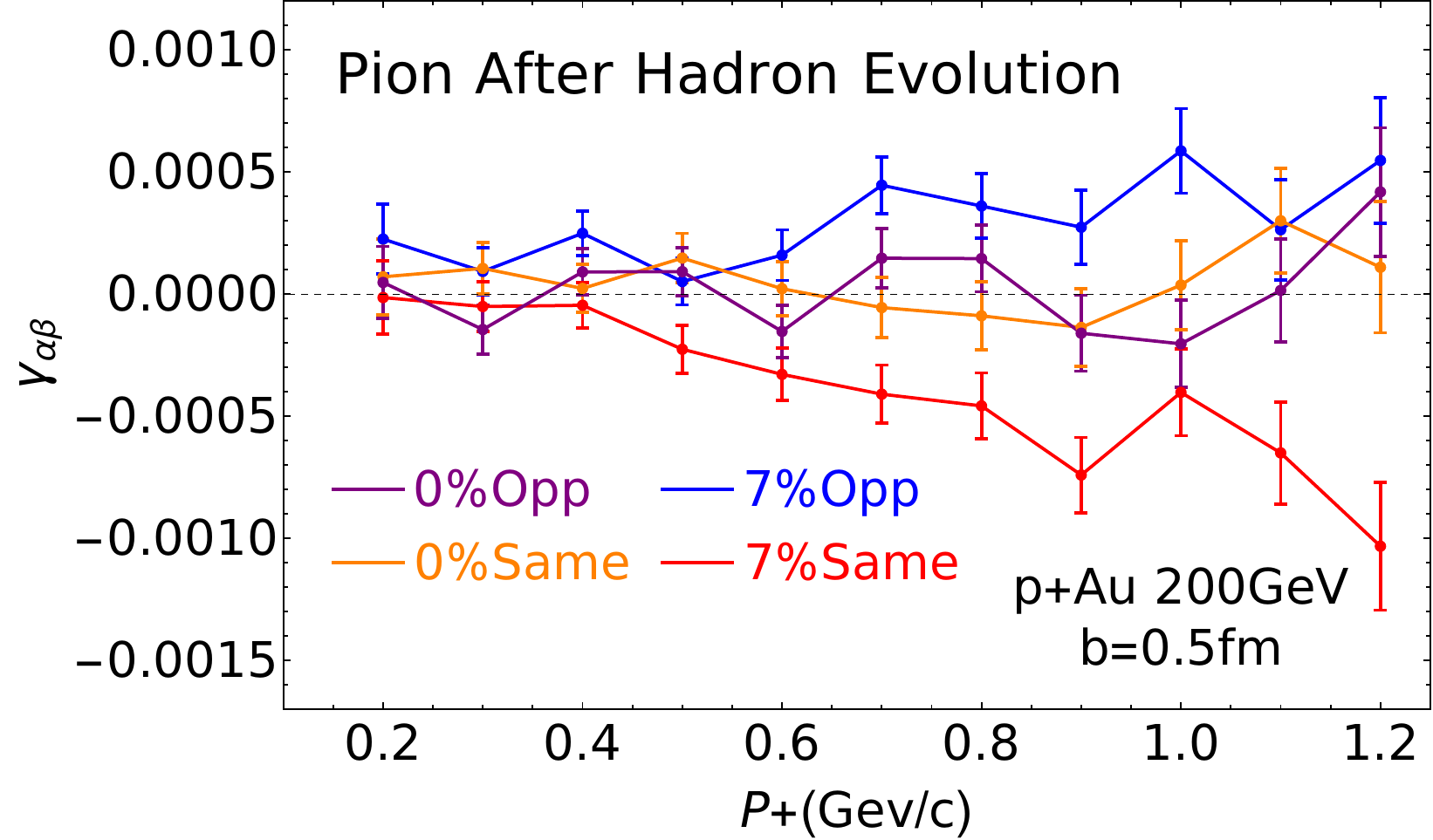}
	\caption{$\gamma_{\alpha\beta}$ as a function of  $P_+$ for $p+Au$ collisions  after hadron level evolution.  Results of  Percentage 7\%  and 0\%  of initial charge separation are shown for comparison. }
	\label{fig:amptpplus}
\end{figure}

The results of   $\gamma_{\alpha\beta}$ at final hadron level as function of $P_+$ are shown  in Fig.~\ref{fig:amptpplus}. These  results with 7\% are similar with the experiment data of $Au+Au$ shown in Fig.~\ref{fig:AA}.  A significant $P_+$ dependent results for both   $\gamma_{OS}$  and   $\gamma_{SS}$ are seen.  In the range of small $p_+$, both of them are depressed very much. The reason for these depression is that the number of quarks with small $P_T$ is much more than number of quarks with large $P_T$. So the quarks with small $p_T$ suffer much more cascade collision, and lose azimuthal correlation much more. 

However, the results show a charge sign dependent clearly. Only $30\%$   values of $\gamma_{OS}$ are survived, while about $50\%$ to $90\%$ values of $\gamma_{SS}$ are transferred  into final hadron level in the range of   $P_+ >0.6$ GeV.  The reason for this phenomenon is perhaps related with the local momentum conservation in resonance decay process.  
In experiment, $\Delta\gamma$ is defined as Eq.~\ref{Eq-Delta-gamma1} to remove this effects.

Also, the results of  0\% are shown in Fig.~\ref{fig:zpcpplus} and Fig.~\ref{fig:amptpplus}. As we expected, they are zero in whole $P_+$ range. These results confirm that the parton evolution and hadron evolution will not bring any additional charge separation caused by background effects in small collision. 
%

\begin{figure}[hb]
	\centering
	\includegraphics[width=0.85\linewidth,clip]{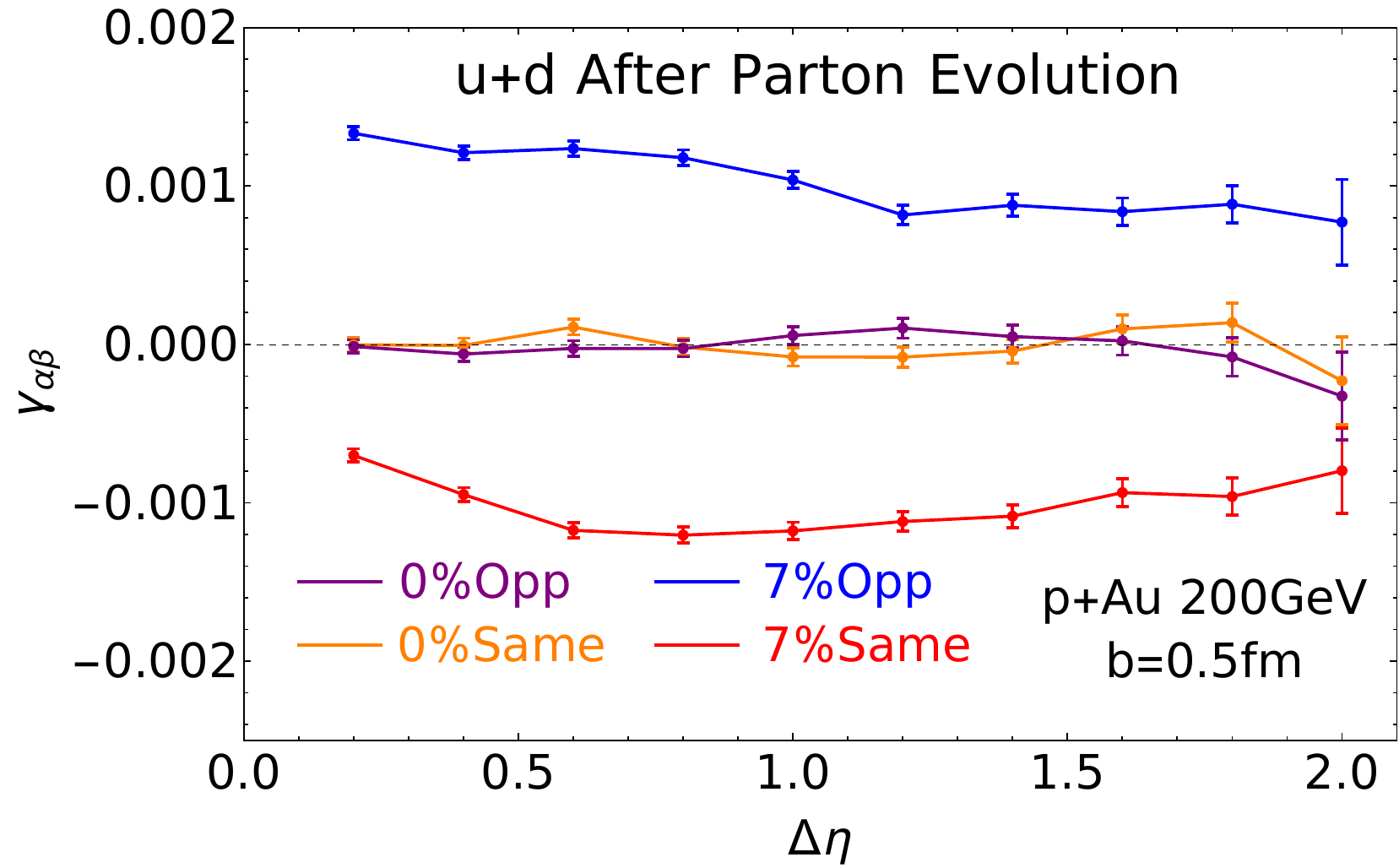}
	\caption{$\gamma_{\alpha\beta}$ as a function of  $\Delta\eta$ for $p+Au$ collisions  after ZPC parton level evolution.  Results of  Percentage 7\%  and 0\%  of initial charge separation are shown for comparison. } 
	\label{fig:zpceta}
\end{figure}

\begin{figure}
	\centering
	\includegraphics[width=0.85\linewidth,clip]{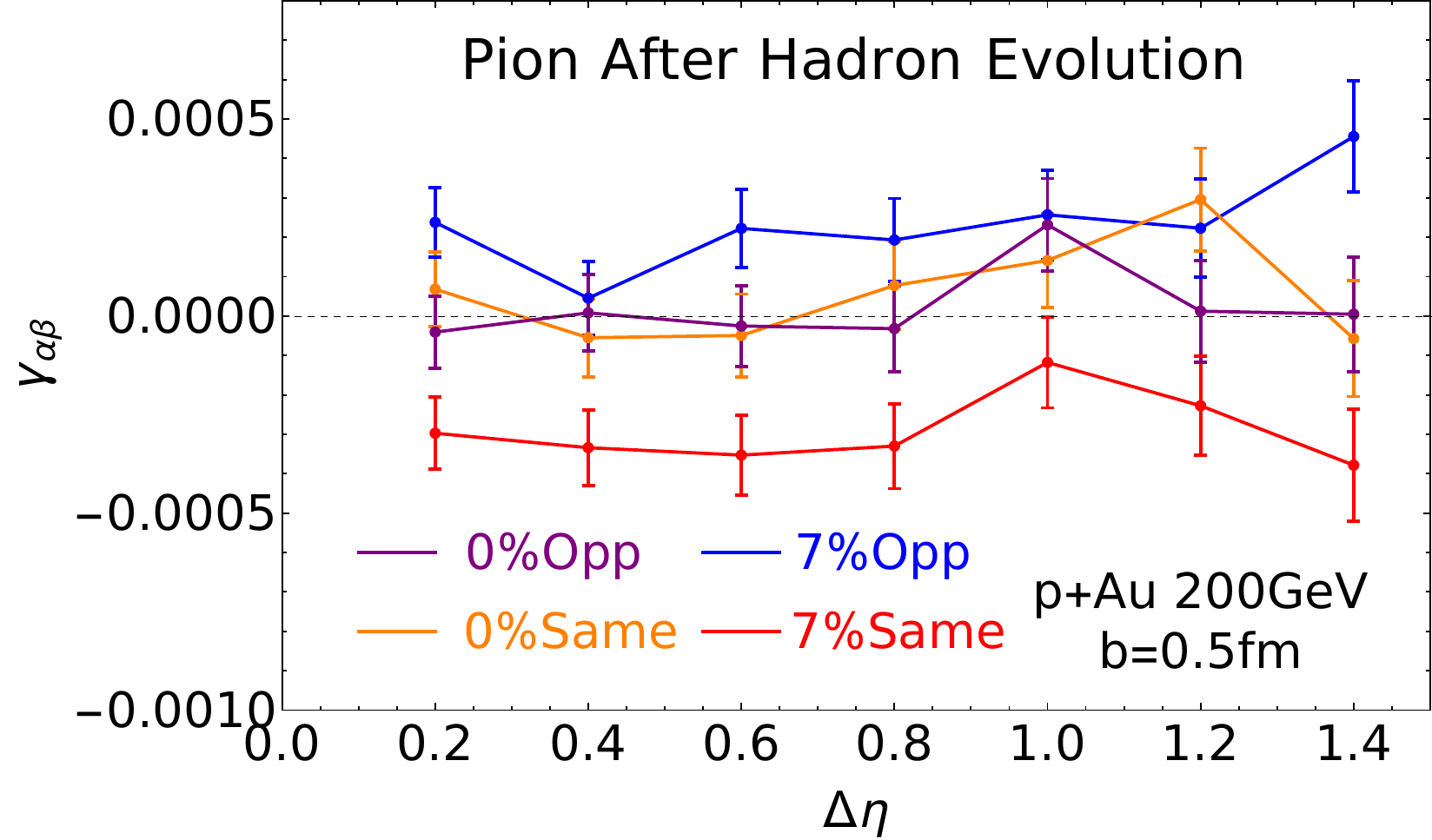}
	\caption{$\gamma_{\alpha\beta}$ as a function of  $\Delta\eta$ for $p+Au$ collisions  after hadron level evolution.  Results of  Percentage 7\%  and 0\%  of initial charge separation are shown for comparison. } 
	\label{fig:ampteta}
\end{figure}

Finally, we check  the correlator $\gamma_{\alpha\beta}$ as function of another kinematic variable $\Delta\eta$ for both parton level and hadron level. $\Delta\eta$  is  defined as ${\Delta\eta}={\left|{\eta_\alpha}-{\eta_\beta}\right|}$ shown in Fig.~\ref{fig:zpceta} and Fig.~\ref{fig:ampteta}. Impact parameter b is set as 0.5 fm in these calculation. 

The results of   $\gamma_{\alpha\beta}$  show us  the value about $\pm 0.001$ are  remained after parton level evolution. And hadron level evolution depress   $\gamma_{OS}$  more than  $\gamma_{SS}$ as we have already seen before. In details, only 15\% of $\gamma_{OS}$ are survived after  hadron level evolution, this ratio for $\gamma_{SS}$ is about  30\%. Unlike  obvious dependence on $P_+$  in Fig.~\ref{fig:amptpplus}, the dependence of $\gamma_{\alpha\beta}$ on $\Delta\eta$ in Fig.~\ref{fig:ampteta} are  negligible. This means a strong longitudinal correlation existing in correlator  $\gamma_{\alpha\beta}$.
These results are quite different with what in $Au+Au$ observed in paper ~\cite{Ma:2011uma} again. The results in that paper  show a strong dependence on the $\Delta\eta$ even with 0\% interchange percentage, indicating the strong bias of near rapidity pairs of $\gamma_{\alpha\beta}$ in $Au+Au$. The difference between $p+Au$ and $Au+Au$ could be explained as the longer lasting time of QGP,  therefor stronger longitudinal correlation of string is broken in heavy ion collisions.

\begin{figure}[hb]
	\centering
	\includegraphics[width=0.85\linewidth,clip]{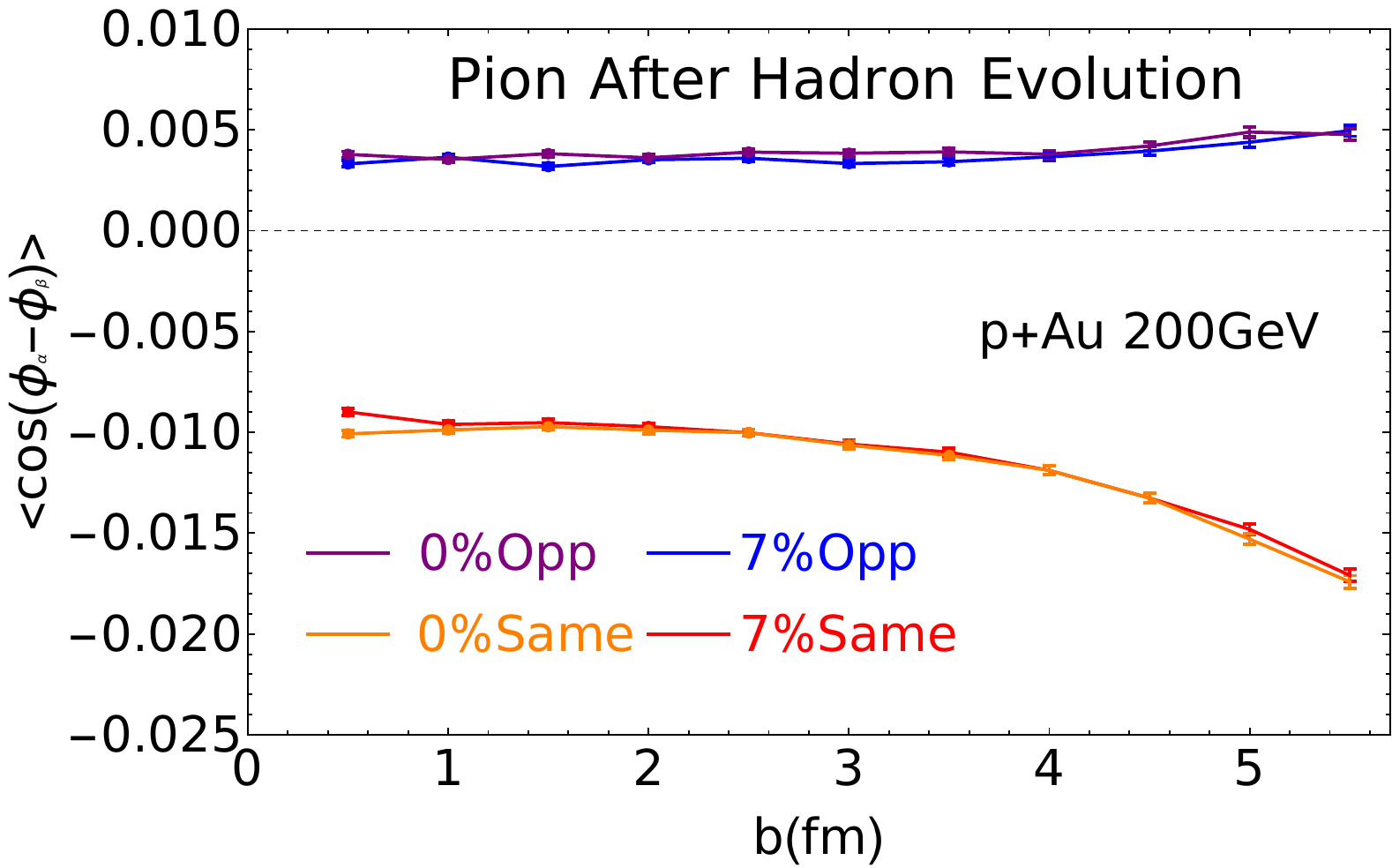}
	\caption{$\tilde{\gamma}_{\alpha\beta}$ as a function of  $b$ for $p+Au$ collisions  after hadron level evolution.  Results of  Percentage 7\%  and 0\%  of initial charge separation are shown for comparison.}
	\label{fig:AreduceB}
\end{figure}

\begin{figure}[ht]
	\centering
	\includegraphics[width=0.85\linewidth,clip]{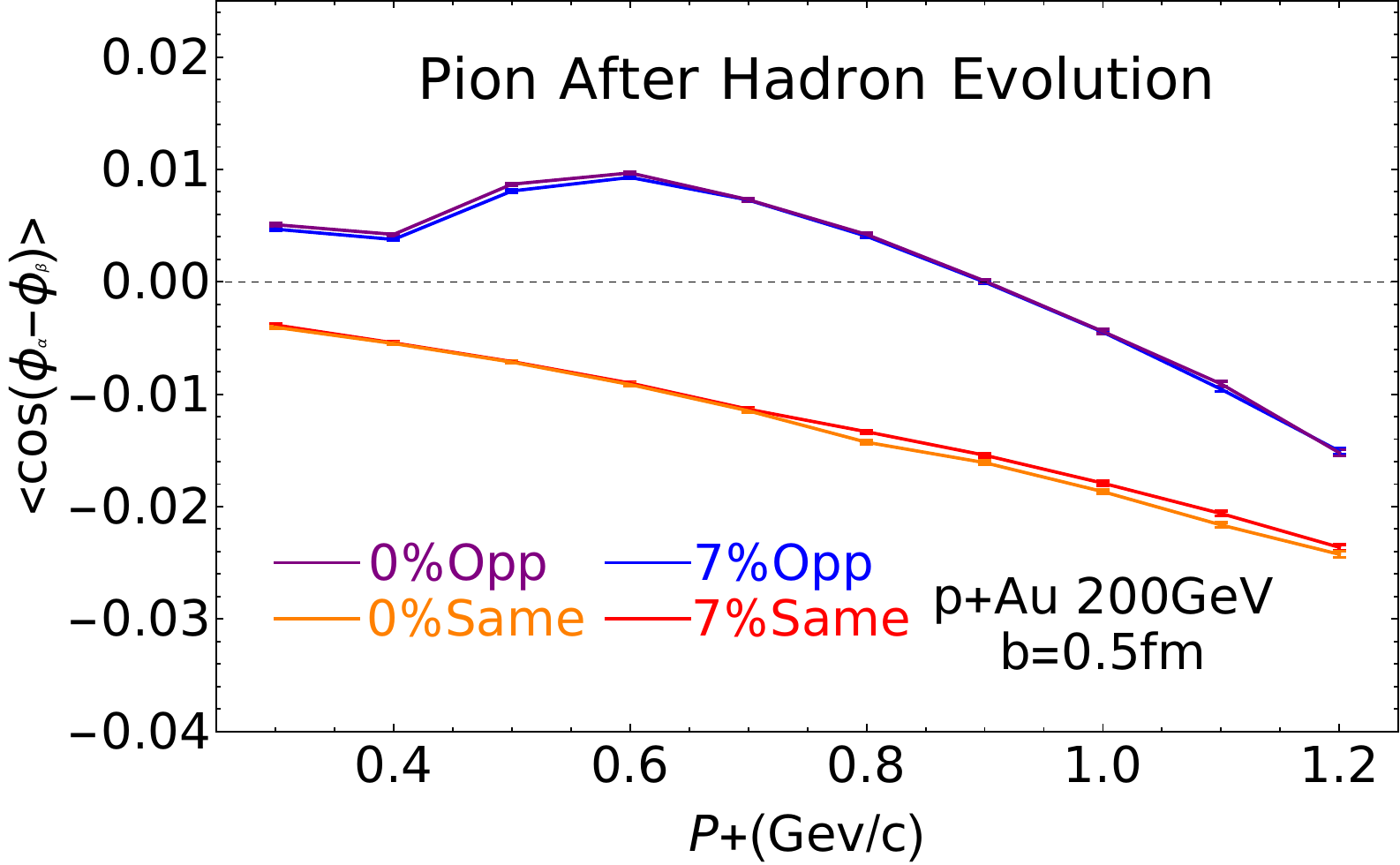}
	\caption{$\tilde{\gamma}_{\alpha\beta}$ as a function of  $P_+$ for $p+Au$ collisions  after hadron level evolution.  Results of  Percentage 7\%  and 0\%  of initial charge separation are shown for comparison.}
	\label{fig:pplusAreduceB}
\end{figure}

The results above are all about the correlator $\gamma_{\alpha\beta}$. And our results of $\gamma_{\alpha\beta}$ have shown effects of backgrounds  are negligible in $p+Au$ collisions. However,  correlator  $\gamma_{\alpha\beta}$ is free of  reaction plane independent backgrounds.  In order to check the effects of all possible backgrounds in $p+Au$ collision, we should defined a new correlator which is sensitive to reaction plane independent backgrounds~\cite{Ma:2011uma}
\begin{equation}
	\tilde{\gamma}_{\alpha\beta}=<\cos(\phi_\alpha-\varphi_\beta)>.
\end{equation}
Our results are shown in Fig.~\ref{fig:AreduceB}  as function of impact parameter b, and Fig.~\ref{fig:pplusAreduceB} as function of $P_+$ correspondingly. 
We can see from these two figures that the  results with percentage 7\% and 0\%  are almost identical. These results prove that the effects of reaction plane independent backgrounds induced by CME mechanism are also negligible in $p+Au$ collision.

\section{Summary \label{sec:Summary}}

In summary, we have studied the effects of parton level evolution and hadron level evolution on the charge separation in small system $p+Au$ collision using AMPT model. An initial charge separation is given by interchange momentum of  7\% quarks supposed as the results of electric current induced by CME. As we have already observed, the effect of evolution can reduce charge separation significantly. But there is still a mount of charge separation signals survived into final hadron system. The results of percentage 0\%  are also prepared as a comparison.  Since there is no CME contribution in the calculation of 0\%, the observed signal should come from background contribution completely. So the zero results means the background has no contribution to correlator $\Delta\gamma$ in $p+Au$ collision. As a conclusion, even though the contribution of CME $\Delta\gamma_\mathrm{CME}$ in $p+Au$ could be a little smaller than that in $Au+Au$, small collision is still a good place to check CME as we proposed in  ~\cite{Zhang:2021jrc,Wu:2024vcd}.
In our future research, we plan to extend our calculation framework to that more closely physical realities. We will set the exchange percentage proportional to the magnetic field strength event by event, and ensure that the direction of exchanged momentum aligns with the magnetic field direction. This will allow us to more realistically check whether small collision systems serve as good method for clarify the contribution of CME effects.

We were supported by the NSFC under Projects No. 12075094. The computation is completed in the HPC Platform of Huazhong University of Science and Technology.
\bibliographystyle{apsrev4-2}
\bibliography{HSCS}

\end{document}